\documentclass[fleqn,10pt]{wlscirep}
\usepackage[utf8]{inputenc}
\usepackage[T1]{fontenc}
\usepackage{siunitx}
\usepackage{bm}

\usepackage{todonotes}
\usepackage{subcaption}

\title{Volume–Surface–Wire Integral Equations for EEG Source Imaging}

\author[1]{Paolo~Ricci}
\author[1]{Maxime~Monin}
\author[1]{Alessandro~Mascherin}
\author[2]{Adrien~Merlini}
\author[1,*]{Francesco~P.~Andriulli}
\affil[1]{Politecnico di Torino, Department of Electronics and Telecommunications, Turin, 10129, Italy}
\affil[2]{IMT Atlantique, Microwave Department, Brest, 29238, France}

\affil[*]{francesco.andriulli@polito.it}

\begin{abstract}
  In EEG source imaging, the precision of the imaging of the brain activity depends on the accuracy of the forward head model which is, in turn, affected by the representation of tissue conductivity, including anisotropic compartments such as the skull and white matter.
  We introduce an EEG source imaging framework based on a hybrid volume-surface-wire integral formulation, enabling tract-aware modeling of anisotropic white matter contributions without a full volumetric head discretization.
  The proposed framework is assessed on a realistic MRI-derived head anatomy, showing close agreement with FEM reference solutions while avoiding full-volume meshing and providing a natural representation of white matter fiber tracts.
  We then integrate this model into a real-time pipeline that couples EEG processing with immersive visualization, enabling synchronized inspection of scalp potentials, cortical sources, and white matter fiber tracts-related quantities. The results support the proposed formulation as a tract-aware alternative to canonical formulations, combining robust localization performance with interactive visualization capabilities.
\end{abstract}
\begin{document}

\flushbottom
\maketitle
%
%
\thispagestyle{empty}

\section*{Introduction}

The study of the electrophysiological activity of the human brain is a cornerstone for a plethora of fields of study, such as clinical neurophysiology, cognitive neuroscience, and neurotechnology, in which a key aspect lies in the reconstruction of neural activity from non-invasive measurements in a form that supports both quantitative analysis and interactive use.
Among non-invasive modalities, electroencephalography (EEG) is especially attractive because it is portable, relatively inexpensive, and provides a temporal resolution in the order of milliseconds.
These characteristics make EEG suitable for both quantitative source imaging and interactive closed-loop settings in which rapid interpretation of neural activity is required~\cite{mishra2014closedloop}.
However, EEG source imaging is intrinsically challenging since the cortical activity must be inferred from scalp potentials by solving an ill-posed inverse problem~\cite{grechReviewSolvingInverse2008a,tait2021systematic}.
In addition to the intrinsic challenges encountered when solving this inverse problem, its setup often requires the use of a forward model that describes how neural sources generate the measured EEG signals.
The forward problem is typically solved multiple times to build a lead-field matrix, which maps source activity to electrode potentials and encodes the effects of head geometry and medium.
Therefore, the assumptions made in the forward model, including those concerning tissue conductivity, can affect the reconstructed sources~\cite{becker2015brainsource,grechReviewSolvingInverse2008a}.
This is especially relevant for tissues such as the skull and white matter, whose inhomogeneous and anisotropic properties have been shown to influence lead-field matrices~\cite{vorwerk2014guidelineb,wolters2006influence,gullmar2010influencea} and, consequently, to affect the accuracy of the imaging process. Representing these properties, however, can be difficult and computationally demanding.

The impact of the numerical method used to solve the forward problem on the  solution of the overall EEG source imaging problem has been analyzed in several studies, and notably involved the finite element method (FEM) and the boundary element method (BEM).
Compared to FEM, BEM approaches discretize tissue interfaces rather than the full head volume and, when combined with fast solution strategies~\cite{kybic2005fastb,baronio_fast_2022}, can lead to reduced memory requirements and computational costs~\cite{meijs1989numerical}.
However, classical BEM formulations generally assume piecewise-homogeneous isotropic compartments, which limits their ability to represent anisotropic conductivity profiles in tissues such as the skull and the white matter.
It is therefore desirable to retain the computational advantages of BEM while extending its applicability to inhomogeneous and anisotropic models.
A possible approach to obtain this is to rely on hybrid integral equation formulations~\cite{monin2020hybridb}, which build on BEM schemes while extending them beyond purely surface-based models through the inclusion of volume and wire integral equations.
In this context, a forward model that explicitly takes into account the white matter fiber tracts is of particular interest, since it provides a practical way to incorporate anatomically structured anisotropic effects while preserving computational advantages typical of BEM schemes; this direction is further supported by fast surface-wire solution strategies with quasi-linear complexity~\cite{baronio_fast_2022}.

Tract-aware anisotropic forward models provide quantities that are not usually available in conventional EEG source imaging pipelines, such as secondary currents associated with white matter fiber representations.
Unlike scalp potentials or cortical source amplitudes, these quantities are distributed along elongated three-dimensional tracts embedded within the head volume.
They are therefore difficult to interpret when inspected only as numerical values or static plots.
In this setting, interactive visualization is useful because it places scalp potentials, reconstructed cortical activity, and tract-related quantities in the same anatomical frame, allowing their spatial relationships and temporal evolution to be inspected jointly.

Real-time VR systems for EEG visualization, intended here as systems in which EEG-derived quantities are continuously updated and displayed in an immersive three-dimensional environment, have already been explored in neurofeedback, rehabilitation, and closed-loop BCI settings~\cite{mishra2014closedloop,li2024combining,delvecchio2024introducing}.
In EEG source imaging, this type of visualization is particularly relevant because the data are simultaneously multichannel, time-varying, and constrained by the subject anatomy.
The inclusion of white matter fiber tracts further increases this spatial complexity, since the quantities of interest are no longer limited to electrode positions and cortical source locations, but also extend to tract-wise variables distributed inside the head model.

White matter fiber tracts are typically obtained by means of diffusion tensor imaging (DTI) tractography, and multimodal studies have often associated these tracts with functional interactions between the cortical regions they connect.
In such cases, however, tract involvement is typically inferred from source activity near tract endpoints or from inter-regional connectivity measures, rather than from electrophysiological quantities explicitly estimated along the tracts themselves~\cite{zhang2022quantitative,basile2022white,deslauriers-gauthier2019white}.
This motivates the integration of the considered forward model with immersive visualization: the hybrid volume-surface-wire formulation provides tract-related secondary-current estimates, while the visualization environment makes these estimates directly inspectable together with scalp and cortical activity.

In this study, we introduce a tract-aware EEG source imaging framework based on the hybrid volume–surface–wire formulation of Monin et al.~\cite{monin2020hybridb}.
First, we assess the framework on realistic head models derived from the IIT Human Brain Atlas v.5.0~\cite{zhang2018evaluation, qi2021regionconnect} by comparing both forward solutions and inverse localization results with an anisotropic FEM reference constructed from the same anatomical data, with matched tissue compartments and conductivity assumptions wherever the two discretization strategies allow.
Second, we exploit the same wire-aware model to build a real-time VR application in which users can interactively inspect three coupled information layers: measured EEG signals, primary currents reconstructed on the cortical surface, and secondary currents along white matter fiber tracts.
This twofold contribution is not intended directly as a method for connectivity estimation.
Rather, it provides a modeling and visualization framework in which tract-associated electrophysiological effects can be inspected together with source estimates.
By explicitly representing tract-related information alongside cortical activity, the platform may support a more physically informed exploration of brain dynamics and offer a complementary perspective to an analysis based only on the correlation between activity in separated brain regions.

The paper is organized as follows.
The \emph{Results} section evaluates the proposed wire-aware EEG source imaging framework, comparing its performance against an anisotropic FEM reference, and presents the real-time visualization outcomes. The \emph{Discussion} section analyzes the results obtained and their implications, and the \emph{Methods} section details the modeling procedures and computational workflow.

\section*{Results}

This section is organized in two sequential stages.
First, we evaluate the proposed source imaging framework by comparing its forward model with that obtained from a state-of-the-art anisotropic FEM pipeline, and by extending the comparison to the inverse problem.
Second, we integrate the same formulation into a real-time pipeline to determine which information can be inspected interactively, with specific attention to white matter fiber tracts.

\subsection*{Formulation assessment}

\begin{figure}
  \centering
  \includegraphics[width=0.92\linewidth]{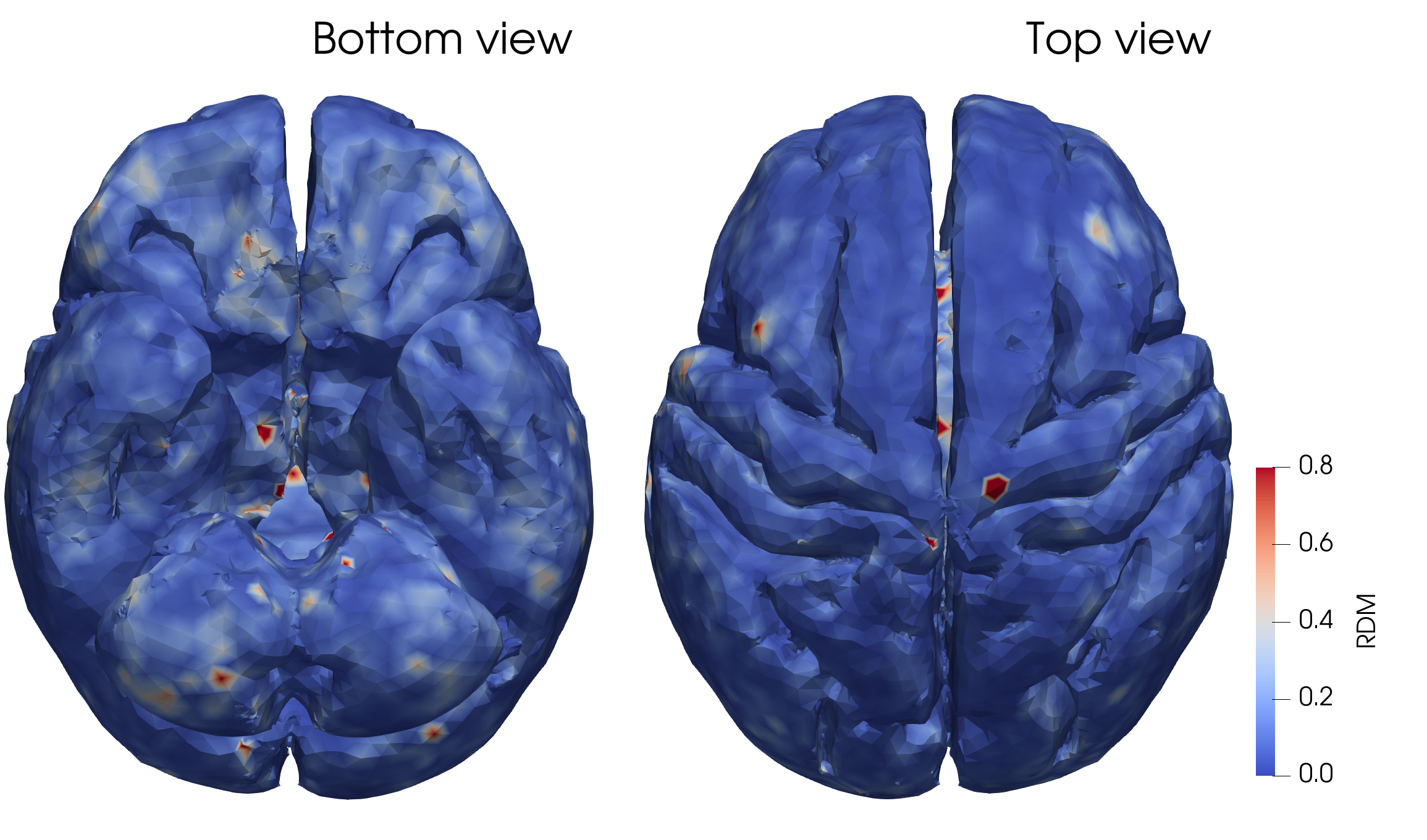}
  \caption{Relative difference measure (RDM) map between the potential at the electrodes predicted by the hybrid volume-surface-wire formulation and by the anisotropic FEM reference. For each unit-amplitude equivalent current dipole placed on the cortical surface, the RDM is computed from the two normalized electrode-potential vectors and assigned to the corresponding source vertex.}
  \label{fig:rdm_bem_fem}
\end{figure}

\begin{figure}
\centering
\begin{subfigure}{0.98\linewidth}
\includegraphics[width=0.92\linewidth]{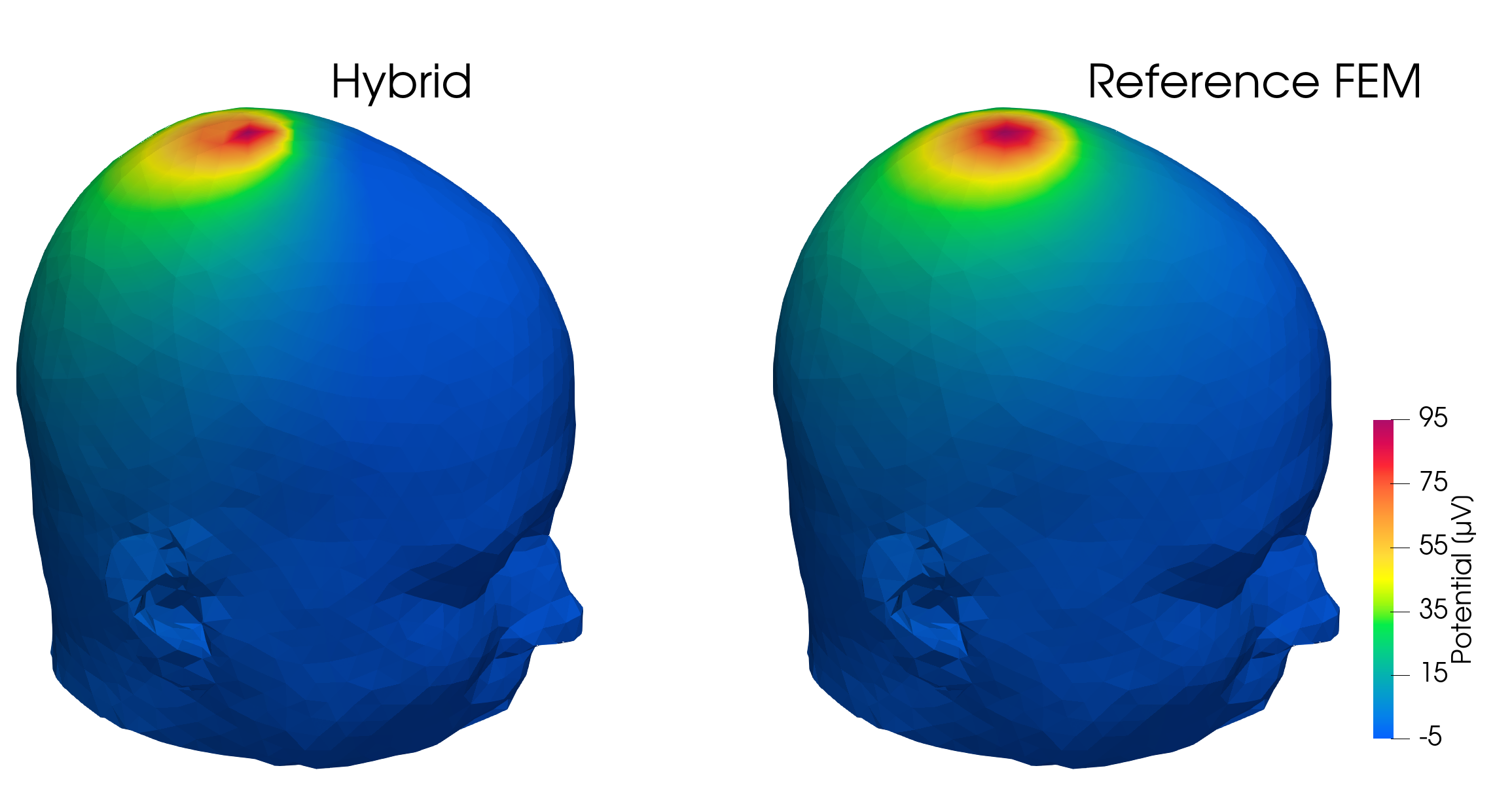}
\caption{High-RDM case.}
\label{fig:high_rdm}
\end{subfigure}  

\begin{subfigure}{0.98\linewidth}
\includegraphics[width=0.92\linewidth]{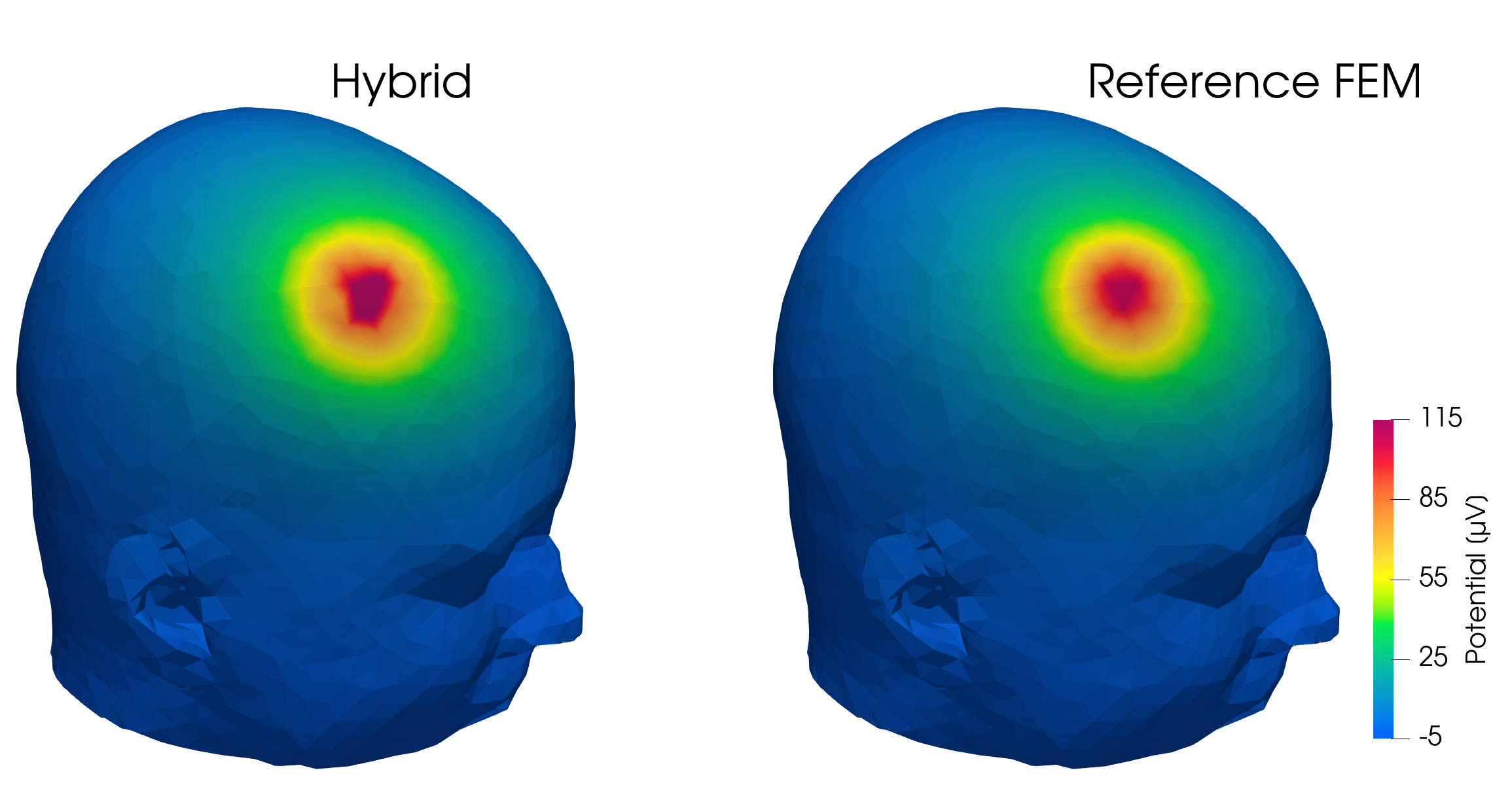}
\caption{Low-RDM case.}
\label{fig:pot_low_rdm}
\end{subfigure} 

\caption{Scalp potential distributions computed with the hybrid volume-surface-wire formulation (left) and with the anisotropic FEM reference (right) for two dipoles: (a) high-RDM case; (b) low-RDM case.} \label{fig:potential_examples}
\end{figure}

The assessment of the proposed formulation is conducted in two steps.
Initially, we compare the forward solutions produced by the hybrid volume-surface-wire formulation and by the anisotropic FEM reference.
For each source location considered, both models are used to compute the vector of scalp potentials at the EEG electrodes.
The discrepancy between the two vectors of potentials is then quantified through the relative difference measure (RDM), which evaluates the difference between the normalized potentials obtained with the two forward models
\begin{equation}
  \mathrm{RDM}_k =
  \left\|
  \frac{\mathbf v^{\mathrm{hyb}}_k}{\|\mathbf v^{\mathrm{hyb}}_k\|_2}
  -
  \frac{\mathbf v^{\mathrm{FEM}}_k}{\|\mathbf v^{\mathrm{FEM}}_k\|_2}
  \right\|_2\,,
\end{equation}
where $\mathbf v^{\mathrm{hyb}}_k$ and $\mathbf v^{\mathrm{FEM}}_k$ are the vectors of potentials generated by the $k$-th source with the hybrid volume-surface-wire and FEM formulation, respectively.
The sources are modeled as equivalent current dipoles located at the vertices of a surface mesh representing gray matter.
Each equivalent current dipole approximates the primary current generated by the coherent activation of a population of pyramidal cells~\cite{baillet_electromagnetic_2001}, which are normal to the cortical surface.
For each simulation, one unit-amplitude dipole is activated at a time.
The corresponding RDM value is then assigned to the vertex where that dipole is located, producing a cortical map that shows, for each candidate source position, how much the potentials predicted by the proposed model differ from the FEM reference.
The resulting map in Fig.~\ref{fig:rdm_bem_fem} shows that the largest discrepancies are mostly located in the inferior region of the head model, whereas superior and lateral regions exhibit a good match.
To show how these discrepancies affect the predicted scalp potentials, Fig.~\ref{fig:potential_examples} shows the potential distributions computed with the two models for two dipoles associated with one of the highest and one of the lowest RDM values, respectively.

In the second step, the impact of the choice of the forward solver on the inverse solution is studied by evaluating the localization accuracy for synthetic EEG data generated with the anisotropic FEM model.
To emulate realistic source-placement uncertainty and to avoid an exact correspondence between the position of the sources used to generate the synthetic data and that of the sources used for inversion, we apply a source-jittering procedure following the strategy described by Gullmar et al.~\cite{gullmar2010influencea}: for each nominal cortical source, the effective source position is displaced with a random perturbation in the range $\SIrange{0.5}{2.5}{mm}$ (and with a mean of $\SI{1.5}{mm}$), and making sure that the sources remain within the admissible head region, before solving the forward problem.
The perturbed source is then used to generate through FEM the synthetic EEG readings. These synthetic EEG measurements are then mapped back to brain activity using both the lead-field obtained with the proposed formulation and once with the (unperturbed) anisotropic FEM reference lead-field.
Each inversion produces a reconstructed source distribution over the cortical source mesh. The accuracy of these solutions is quantified in terms of localization error (LE), defined as the Euclidean distance between the location of the maximum reconstructed activity $\bm{r}_{\mathrm{peak}}$ and the true source location $\bm{r}_{\mathrm{true}}$
\begin{equation}
\mathrm{LE} = \|\bm{r}_{\mathrm{peak}} - \bm{r}_{\mathrm{true}}\|_2\,,
\end{equation}
which is a standard metric in EEG source imaging validation~\cite{grechReviewSolvingInverse2008a,becker2015brainsource}.
This procedure produces one LE value for each tested nominal source vertex.
Each value is computed from the reconstruction obtained from the corresponding perturbed FEM source and is then assigned to the nominal vertex from which the perturbation was generated.
Repeating the procedure over all sources yields cortical LE maps that report localization accuracy as a function of source position.

Two separate noise conditions are evaluated: a noise-free case and a noisy case with white Gaussian noise imposed to the EEG readings so that the signal-to-noise ratio is $\mathrm{SNR} = \SI{5}{dB}$, in line with common EEG protocols for the validation of inverse solutions~\cite{grechReviewSolvingInverse2008a}.
In both cases, the weighted minimum-norm estimate (wMNE)~\cite{pascual-marqui1999review} and standardized low-resolution brain electromagnetic tomography (sLORETA)~\cite{pascual-marqui2002standardized} are used to solve the inverse problem. 

\begin{figure}
\centering
\begin{subfigure}{0.98\linewidth}
  \centering
  \includegraphics[width=0.92\linewidth]{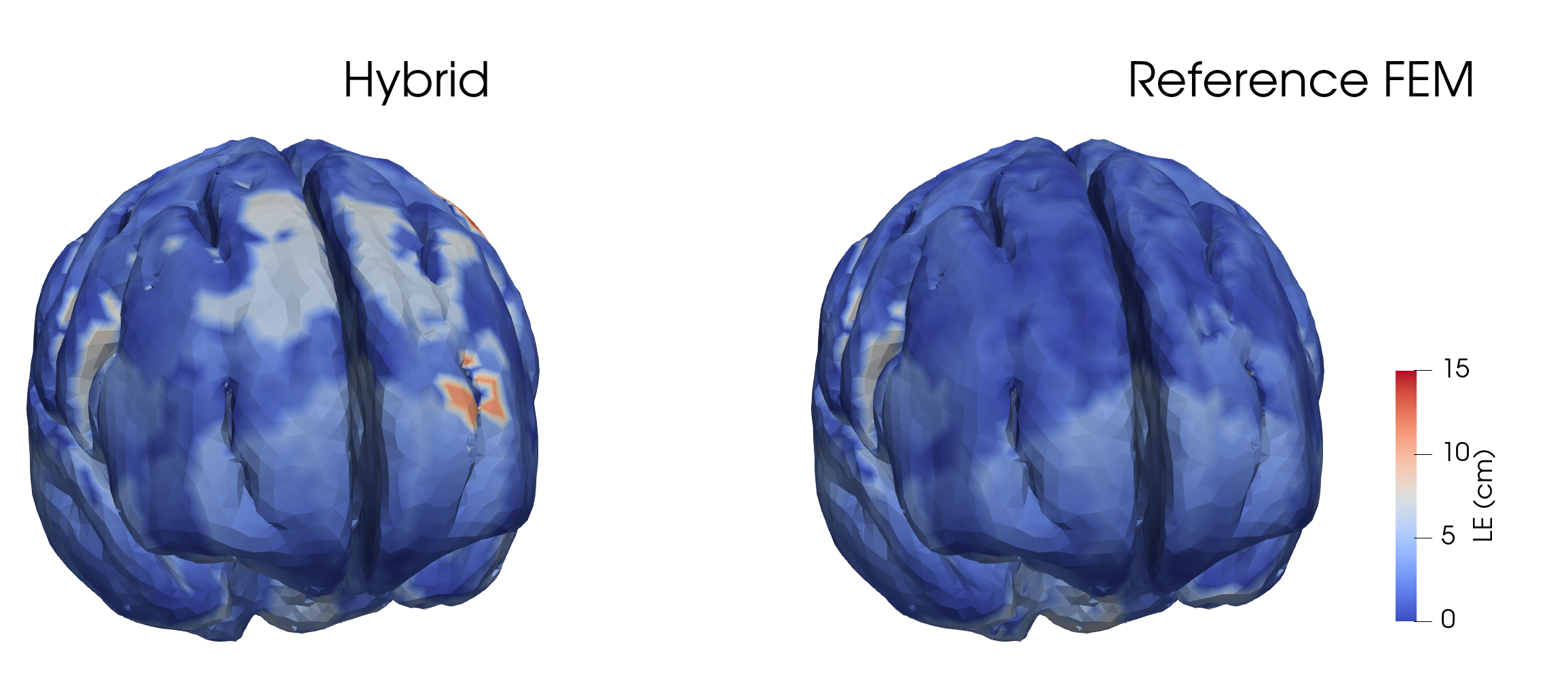}
  \caption{wMNE}
  \label{fig:loc_err_wmne_nonoise}
\end{subfigure}

\begin{subfigure}{0.98\linewidth}
  \centering
  \includegraphics[width=0.92\linewidth]{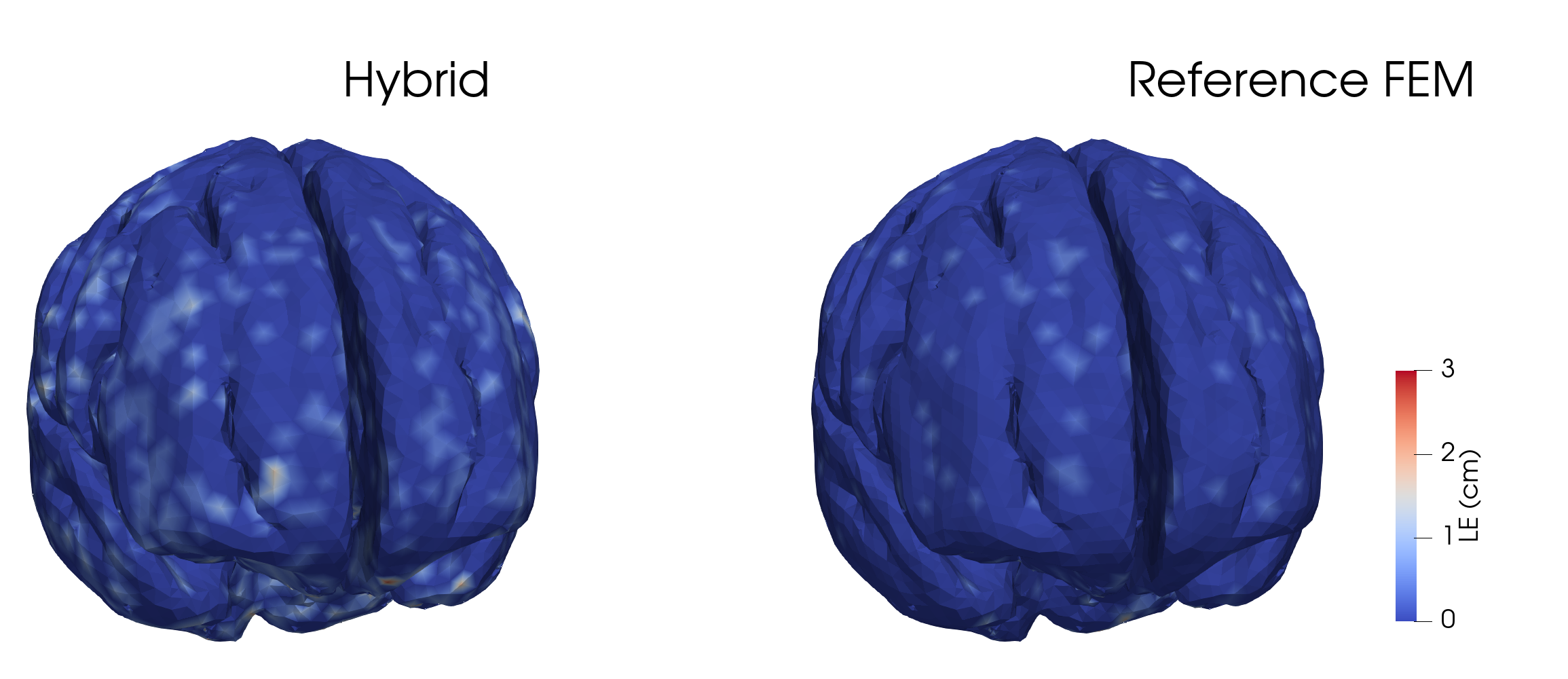}
  \caption{sLORETA}
  \label{fig:loc_err_sloreta_nonoise}
\end{subfigure}
\caption{Localization error maps using the hybrid volume-surface-wire formulation (left) and the FEM reference (right) in the noise-free setting: (a) wMNE; (b) sLORETA.}
\label{fig:noiseless}
\end{figure}

\begin{figure}
\centering
\begin{subfigure}{0.98\linewidth}
  \centering
  \includegraphics[width=0.92\linewidth]{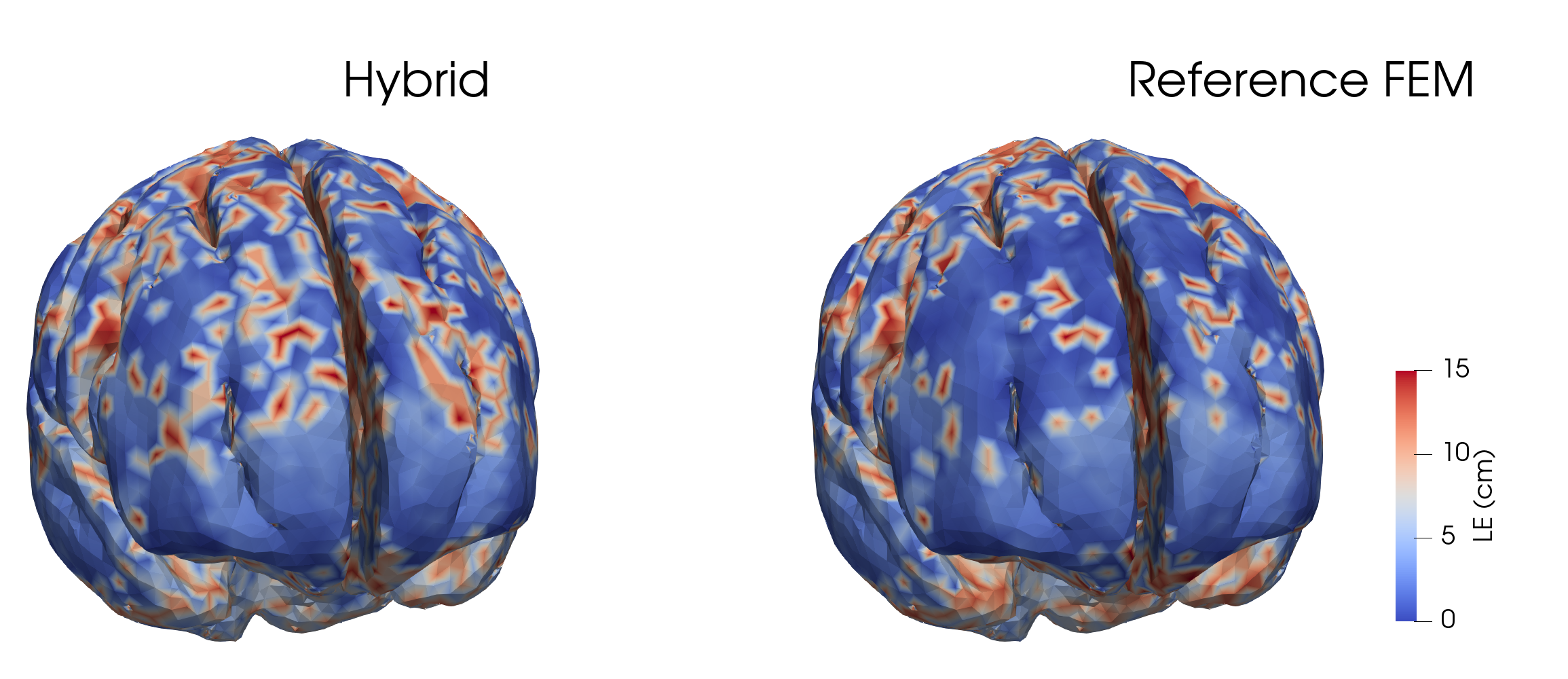}
  \caption{wMNE}
  \label{fig:loc_err_wmne_noise}
\end{subfigure}

\begin{subfigure}{0.98\linewidth}
  \centering
  \includegraphics[width=0.92\linewidth]{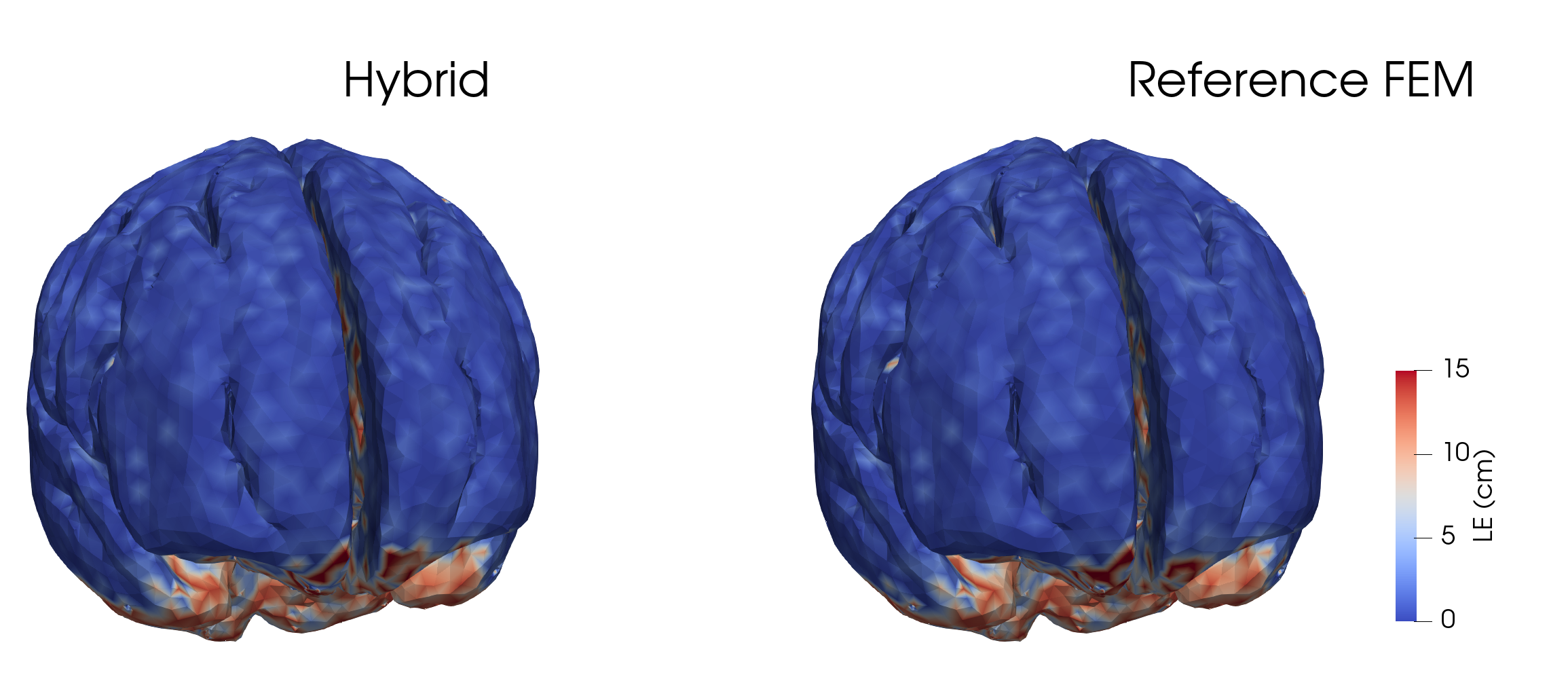}
  \caption{sLORETA}
  \label{fig:loc_err_sloreta_noise}
\end{subfigure}
\caption{Localization error maps using the hybrid volume-surface-wire formulation (left) and the FEM reference (right) in the noisy ($\mathrm{SNR} = \SI{5}{dB}$) setting: (a) wMNE; (b) sLORETA.}
\label{fig:noisy}
\end{figure}

The LE maps in Fig.~\ref{fig:noiseless} and Fig.~\ref{fig:noisy} summarize the inverse comparison performed in the noise-free and in the noisy setting, respectively.
For both wMNE and sLORETA, the LE maps associated with the lead-field obtained with the hybrid volume-surface-wire and with the anisotropic FEM reference are comparable.
Introducing noise increases the localization error for both inverse methods.
Nonetheless, the increase in LE from the noise-free to the noisy condition is comparable for the two formulations.

The agreement between the LE obtained with the two forward formulations is assessed by performing a two one-sided test (TOST)~\cite{lakens_equivalence_2017}.
The test evaluates whether the mean of the paired differences in LE between reconstructions, obtained with the two lead-fields considered, lies within a predefined equivalence interval.
We set the equivalence margin to $\delta = \SI{4}{mm}$, a value below the \SI{5}{mm} spatial-accuracy target suggested for EEG source imaging by Baillet and Garnero~\cite{baillet_bayesian_1997}, and use a significance level of $\alpha = 0.005$.
Statistical equivalence is observed for all four combinations of inverse methods and noise conditions: in each case, the corresponding 99\% confidence interval of the mean of the paired LE differences lies entirely within the equivalence bounds $[-\delta,\delta]$.

\subsection*{Real-time visualization tool}

\begin{figure}
  \centering
  \includegraphics[width=0.92\linewidth]{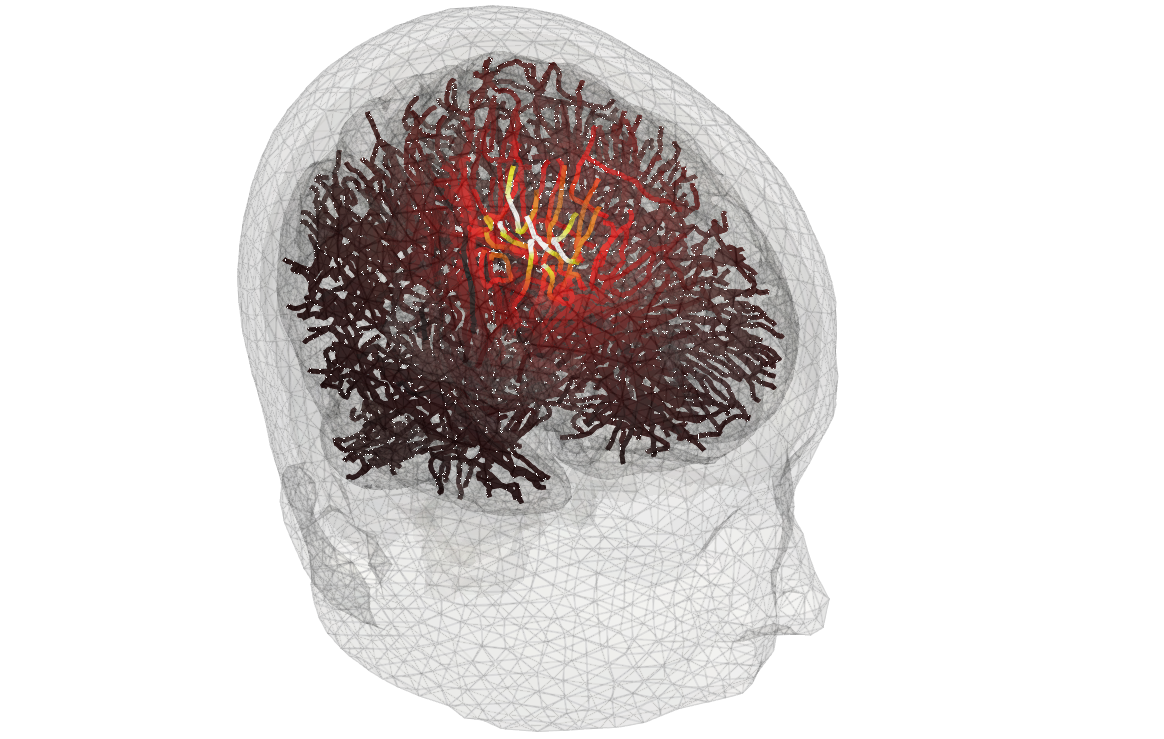}
  \caption{Example of a tract-related quantity provided by the volume-surface-wire formulation. The electric potential generated by a dipole located on the cortical surface is represented along the white matter fiber tracts, showing how spatial information can be extracted directly from the wire component of the formulation.}
  \label{fig:fiber_currents}
\end{figure}

The previous validation results are then exploited in the visualization stage.
In real-time applications, the FEM-based reference provides a strong benchmark for scalp and cortical source imaging but does not natively provide tract-related quantities at interactive rates.
By contrast, the volume-surface-wire formulation directly outputs tract-related variables, as shown for example in Fig.~\ref{fig:fiber_currents}.
This enables synchronized rendering of scalp potentials, cortical activity, and white matter fiber tract activity within the same analysis loop.

\begin{figure}
  \centering
  \includegraphics[width=0.3\linewidth]{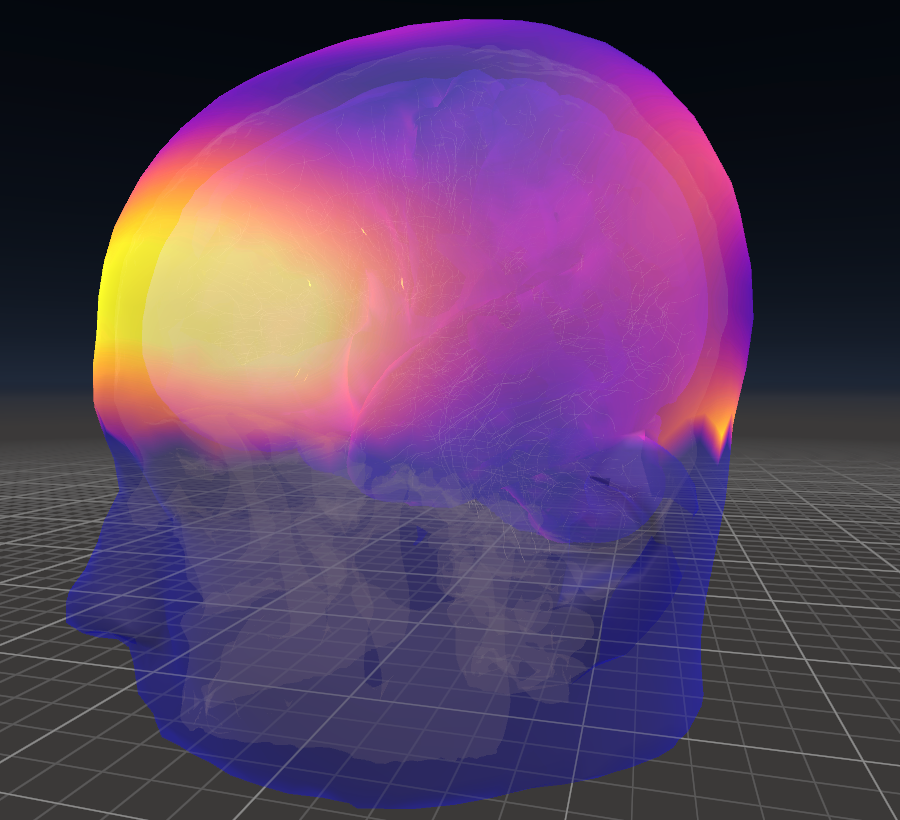}
  \includegraphics[width=0.3\linewidth]{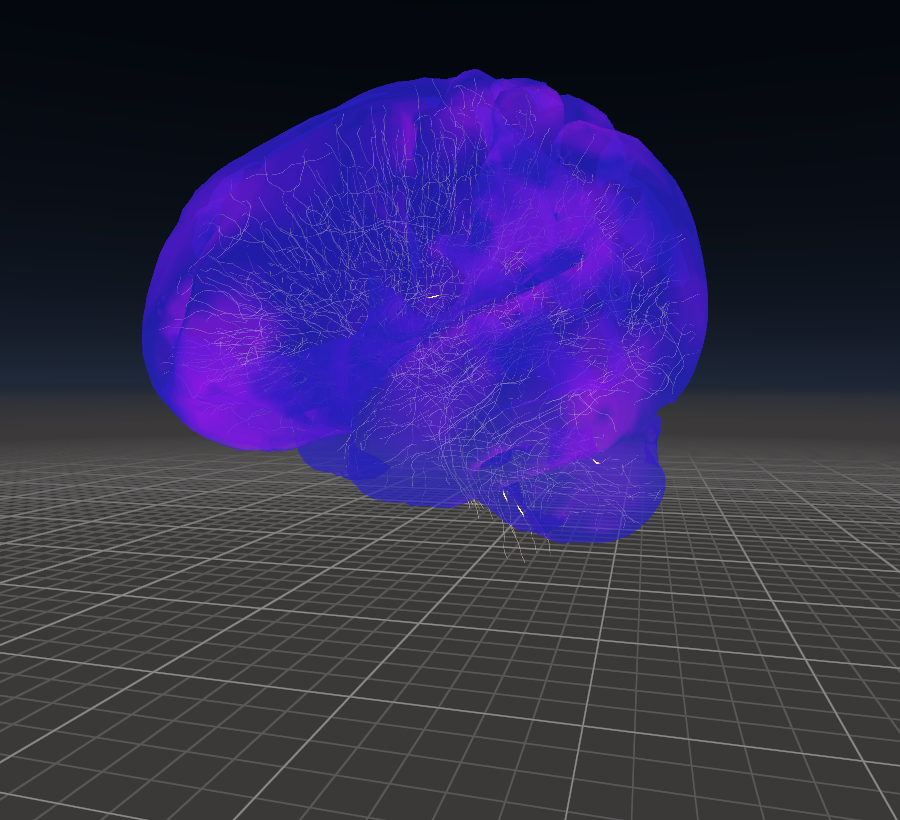}
  \includegraphics[width=0.3\linewidth]{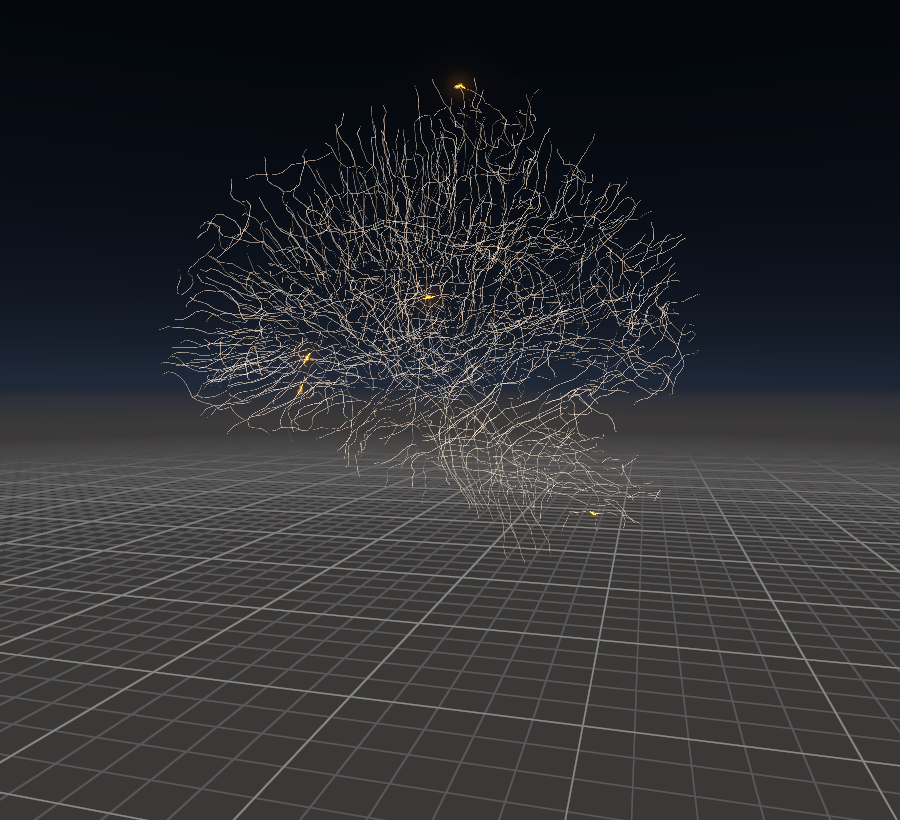}
  \caption{Example of the real-time visualization interface with different anatomical layers enabled: (left) scalp, skull, cortical surface, and white matter fiber tracts; (center) cortical surface and white matter fiber tracts; (right) white matter fiber tracts only.}
  \label{fig:neurosurf}
\end{figure}

The developed tool supports immersive exploration of the full processing chain in real-time.
Incoming EEG data and inverse outputs are mapped onto a multilayer anatomical model (scalp, skull, cortex, and tracts), while the user can navigate the scene and interactively adjust model pose, scale, colormap, and layer opacity.
An example of tool usage is shown in Fig.~\ref{fig:neurosurf}, illustrating how the operator can inspect the correspondence between scalp patterns, reconstructed cortical sources, and tract-related activity.

\section*{Discussion}

The results obtained suggest that the hybrid volume-surface-wire formulation is consistent with the anisotropic FEM reference, with some spatially localized discrepancies.
Moreover, the explicit wire representation of white matter fiber tracts makes that formulation suitable for a real-time visualization pipeline in which scalp potentials, reconstructed cortical activity, and tract-related quantities can be inspected within a common anatomical model.

In the forward comparison, the mismatch with the anisotropic FEM reference remains mainly localized in the inferior region of the head model.
In the inverse setting, both wMNE and sLORETA recover broadly consistent localization patterns across most cortical areas.
Introducing measurement noise increases the LE, as expected, but the degradation from the noise-free to the noisy condition is similar for the two forward formulations.
This agreement is supported by the equivalence analysis, which shows statistical equivalence between the two formulations for all combinations of inverse method and noise condition.
This suggests that replacing the FEM lead-field with that obtained from the hybrid volume-surface-wire formulation does not introduce a practically relevant change in average LE.

A key point for interpretation is that the comparison is asymmetric: synthetic measurements are generated with the (sources-perturbed) anisotropic FEM model and then inverted using both lead-fields, obtained with FEM and with the proposed method.
This setup favors the FEM-based inversion, because one of the inverse models is more closely matched to the model used to generate the data.
For this reason, small differences in localization error should be interpreted while accounting for this model mismatch, rather than being taken alone as evidence of instability of the proposed formulation.
Moreover, the spatial distribution of the differences is coherent with expectations: inferior and sulcal regions are generally more challenging because deeper sources are more dependent on the conductivity assumptions in anatomically complex areas~\cite{wolters2006influence,gullmar2010influencea,dannhauer2010modeling}.

From an application perspective, an advantage of the proposed formulation is that it retains good performances in the EEG source imaging problem while providing tract-related quantities that can be visualized in interactive settings.
This allows the development of a system that is the first of its kind to directly derive white matter involvement from a wire equation solved through a hybrid integral equation formulation.
This approach to the evaluation of the activation of the white matter fiber tracts provides a novel way to visualize the relationship between cortical activation and structural connectivity.
This is particularly relevant as it enables users to inspect scalp potentials, cortical reconstructions, and white matter fiber tracts activity in a synchronized way, which might prove useful in several fields such as neurofeedback and BCI analyses~\cite{mishra2014closedloop, mullen2013realtime, balloni2024brain}.

\section*{Methods}

The proposed framework, available on Zenodo~\cite{ricci_2026_22035770}, is organized into two independent modules: a \emph{data module} and a \emph{visualization module}.
The data module, implemented in Python, is responsible for EEG acquisition, preprocessing, and solving the source imaging problem.
The visualization module, developed in Unity, manages the immersive VR environment and user interaction.
The two modules are connected solely through the exchange of data streams, which are transmitted via the Lab Streaming Layer (LSL) protocol~\cite{kothe2025lab}.
The quantities transmitted include potentials at the scalp level, reconstructed source-level activations, and potential differences between the endpoints of the fiber tracts.
Both modules operate on a common anatomical brain model, which is defined during the setup stage.

In what follows, we first propose a rapid overview of the hybrid volume-surface-wire forward model employed, we then describe the anatomical model, followed by details on the data pipeline from EEG acquisition to the solution of the EEG source imaging problem, and finally explain the visualization and interaction strategies.

\subsection*{Hybrid volume-surface-wire formulation}

A key component of the EEG source imaging problem is the forward model, which is used to build the lead-field matrix that maps dipole sources to electrode potentials.
In this work, we employ a hybrid integral equation formulation that extends the classical surface-based approach by also including one-dimensional white matter fiber tracts modeled by means of a wire equation and an anisotropic skull volume.
The problem setting presented in the following is based on the formulation introduced by Monin et al.~\cite{monin2020hybridb}.

Taking into account a model made of $N$ nested layers, we denote by $\Omega_i$, with $ \Omega = \bigcup_{i=1}^{N} \Omega_i$, the domain of the $i-th$ medium that composes the  model, and by $\Gamma_i$, with $\Gamma = \bigcup_{i=1}^{N} \Gamma_i$, the external contour of $\Omega_i$, i.e., $\Gamma_i = \partial \Omega_i \cap \partial \Omega_{i+1}$.
In this context, solving the forward problem corresponds to determining the electric potential $\phi(\mathbf r)$ generated by a source positioned in $\mathbf r \in \Omega$ that induces a primary current $\mathbf J_p(\mathbf r)$.
In the quasi-static regime, this corresponds to solving Poisson's equation
\begin{equation}
  \nabla \cdot \bigl(\boldsymbol{\sigma}(\mathbf r)\nabla \phi(\mathbf r)\bigr) = \nabla \cdot \mathbf J_p(\mathbf r), \qquad \mathbf r \in \Omega
\end{equation}
setting as boundary conditions
\begin{subequations}
  \begin{align}
    \phi_i^{-}(\mathbf r) &= \phi_i^{+}(\mathbf r), \qquad \mathbf r \in \Gamma_i\,,  \\
    \hat{\mathbf n}(\mathbf r)\cdot \boldsymbol{\sigma}(\mathbf r)\nabla \phi_i^{-}(\mathbf r)  &=  \hat{\mathbf n}(\mathbf r)\cdot \boldsymbol{\sigma}(\mathbf r)\nabla \phi_i^{+}(\mathbf r), \qquad \mathbf r \in \Gamma_{i<N}\,,  \\
    \hat{\mathbf n}(\mathbf r)\cdot \boldsymbol{\sigma}\nabla \phi(\mathbf r) &= 0, \qquad \mathbf r \in \Gamma_N\,,
  \end{align}
\end{subequations}
where $\phi_i^{-}(\mathbf r)$ (respectively $\phi_i^{+}(\mathbf r)$) is the electric potential just inside (outside) $\Gamma_i$ in $\mathbf r$, $\hat{\mathbf n}(\mathbf r)$ is the unit vector normal to the interface and pointing outward, and $\boldsymbol{\sigma}(\mathbf r)$ is the tissue conductivity tensor in $\mathbf r$, which is simplified to $\sigma_i$ where the $i-th$ medium is isotropic and homogeneous.

The surface and volume operators are
\begin{subequations}
  \begin{align}
    (\mathcal S f)(\mathbf r)  &=  \int_{\Gamma} G(\mathbf r,\mathbf r')\,f(\mathbf r')\,d\Gamma',  \qquad  \mathbf r \in \Omega ,  \\
    (\mathcal S_v\mathbf f)(\mathbf r)  &=  \int_{\Omega} G(\mathbf r,\mathbf r')\,\nabla'\cdot \mathbf f(\mathbf r')\,d\Omega',  \qquad  \mathbf r \in \Omega ,  \\
    (\mathcal D^{*} f)(\mathbf r)  &=  \int_{\Gamma}  \hat{\mathbf n}(\mathbf r)\cdot \nabla G(\mathbf r,\mathbf r')\,  f(\mathbf r')\,d\Gamma',  \qquad  \mathbf r \in \Gamma ,  \\
    (\mathcal D_v^{*}\mathbf f)(\mathbf r)  &=  \int_{\Omega}  \hat{\mathbf n}(\mathbf r)\cdot \nabla G(\mathbf r,\mathbf r')\,  \nabla'\cdot \mathbf f(\mathbf r')\,d\Omega',  \qquad  \mathbf r \in \Gamma \,,
  \end{align}
\end{subequations}
where $G(\mathbf r,\mathbf r') = \frac{1}{4\pi \|\mathbf r-\mathbf r'\|}$ is the static Green's function.
Then, we expand surface unknowns on the associated triangle meshes with $N_s$ pyramid functions $\{\psi_n^{(s)}\}_{n=1}^{N_s}$.
For a surface vertex $\mathbf r_n$, let $t_{nkl}$ be any triangle including it, with vertices $\mathbf r_n$, $\mathbf r_k$, and $\mathbf r_l$.
The pyramid function associated with $\mathbf r_n$ is defined as
\begin{equation}
  \psi_n^{(s)}(\mathbf r) =
  \begin{cases}
    \dfrac{
      \left\|(\mathbf r-\mathbf r_l)\times(\mathbf r_k-\mathbf r_l)\right\|
    }{
      \left\|(\mathbf r_n-\mathbf r_l)\times(\mathbf r_k-\mathbf r_l)\right\|
    },
    & \mathbf r \in t_{nkl},\\
    0,
    & \text{otherwise}.
  \end{cases}
\end{equation}

Inside the skull we expand volumetric unknowns with $N_v$ Schaubert-Wilton-Glisson (SWG) basis functions~\cite{schaubert_tetrahedral_1984} $\{\boldsymbol\psi_n^{(v)}\}_{n=1}^{N_v}$ defined over a volumetric discretization made of tetrahedra.
Each SWG function is associated with an internal triangular face $t_n$ shared by two tetrahedra $T_n^+$ and $T_n^-$.
Let $a_n$ be the area of $t_n$, $V_n^\pm$ the volumes of $T_n^\pm$, and $\mathbf r_n^\pm$ the vertices of $T_n^\pm$ opposite to $t_n$.
After fixing the orientation from $T_n^+$ to $T_n^-$, the corresponding SWG basis function is defined as
\begin{equation}
  \boldsymbol\psi_n^{(v)}(\mathbf r) =
  \begin{cases}
    \dfrac{a_n}{3V_n^+}\left(\mathbf r-\mathbf r_n^+\right),
    & \mathbf r \in T_n^+, \\
    \dfrac{a_n}{3V_n^-}\left(\mathbf r_n^- - \mathbf r\right),
    & \mathbf r \in T_n^-, \\
    \mathbf 0,
    & \text{otherwise}.
  \end{cases}
\end{equation}
The choice of $T_n^+$ and $T_n^-$ fixes the sign convention used in the assembly.

White matter fiber tracts are represented, instead, as oriented one-dimensional manifolds.
For a node $\mathbf r_n$ of a fiber, let $\mathbf r_{n-1}$ and $\mathbf r_{n+1}$ be the previous and following nodes along the chosen fiber orientation.
The piecewise linear hat function associated with node $\mathbf r_n$ having support on the segments $s_n^-=[\mathbf r_{n-1},\mathbf r_n]$ and $s_n^+=[\mathbf r_n,\mathbf r_{n+1}]$, is defined as
\begin{equation}
  \boldsymbol\psi_n^{(w)}(\mathbf r) =
  \begin{cases}
    \dfrac{\mathbf r-\mathbf r_{n-1}}
          {\|\mathbf r_n-\mathbf r_{n-1}\|}\,,
    & \mathbf r \in s_n^{-}, \\
    \dfrac{\mathbf r_{n+1}-\mathbf r}
          {\|\mathbf r_{n+1}-\mathbf r_n\|}\,,
    & \mathbf r \in s_n^{+}, \\
    \mathbf 0,
    & \text{otherwise}.
  \end{cases}
\end{equation}

We collect the corresponding coefficients into vectors $\mathsf x_s$ for surfaces, $\mathsf x_v$ for skull volume, and $\mathsf x_w$ for wires.
Anisotropy is then imposed for the fiber tracts and the skull, leveraging the conductivity contrast to the background medium, defined as $\boldsymbol{\chi}_i(\mathbf r)  =  \bigl(\sigma_i \mathbf I - \boldsymbol{\sigma}(\mathbf r)\bigr) \boldsymbol{\sigma}(\mathbf r)^{-1}$.
With these definitions, the discrete system reads
\begin{equation}
  \begin{bmatrix}
    -\mathsf G_{ss} + \mathsf D_{ss}^{*} & \mathsf G_{sv} - \mathsf D_{sv}^{*} & -\mathsf D_{sw}^{*}\\
    -\mathsf S_{vs} & \mathsf G_{vv} + \mathsf S_{vv} & \mathsf S_{vw} \\
    -\mathsf S_{ws} & \mathsf S_{wv} & \mathsf G_{ww} + \mathsf S_{ww}
  \end{bmatrix}
  \begin{bmatrix}
    \mathsf x_s\\
    \mathsf x_v\\
    \mathsf x_w
  \end{bmatrix}
  =
  \begin{bmatrix}
    \mathsf b_s\\
    \mathsf b_v\\
    \mathsf b_w
  \end{bmatrix}\,,
\end{equation}
where the block operators are defined as
\begin{subequations}
  \begin{align}
    (\mathsf G_{ss})_{mn}  &=  \frac{\sigma_m^{-}+\sigma_m^{+}}{2(\sigma_m^{+}-\sigma_m^{-})}  \,  \bigl\langle \psi_m^{(s)},\psi_n^{(s)}\bigr\rangle_{\Gamma},  \\
    (\mathsf D_{ss}^{*})_{mn}  &=  \bigl\langle \psi_m^{(s)}, \mathcal D^{*}\psi_n^{(s)}\bigr\rangle_{\Gamma},  \\
    (\mathsf G_{sv})_{mn}  &=  \left\langle  \psi_m^{(s)},  \frac{1}{2\sigma_m^{-}}\,  \hat{\mathbf n}\cdot  \boldsymbol{\chi}\,\boldsymbol{\psi}_n^{(v)}  \right\rangle_{\Gamma},  \\
    (\mathsf D_{sv}^{*})_{mn}  &=  \left\langle  \psi_m^{(s)},  \frac{1}{\sigma_m^{-}}\,  \mathcal D_v^{*}\bigl(\boldsymbol{\chi}\,\boldsymbol{\psi}_n^{(v)}\bigr)  \right\rangle_{\Gamma},  \\
    (\mathsf D_{sw}^{*})_{mn}  &=  \left\langle  \psi_m^{(s)},  \frac{1}{\sigma_m^{-}}\,  \mathcal D_v^{*}\bigl(\boldsymbol{\chi}\,\boldsymbol{\psi}_n^{(w)}\bigr)  \right\rangle_{\Gamma},  \\
    (\mathsf S_{vs})_{mn}  &=  \bigl\langle  \boldsymbol{\psi}_m^{(v)},  \nabla \mathcal S \,\psi_n^{(s)}  \bigr\rangle_{\Omega},  \\
    (\mathsf G_{vv})_{mn}  &=  \bigl\langle  \boldsymbol{\psi}_m^{(v)},  \boldsymbol{\chi}^{-1}\boldsymbol{\psi}_n^{(v)}  \bigr\rangle_{\Omega},  \\
    (\mathsf S_{vv})_{mn}  &=  \left\langle  \boldsymbol{\psi}_m^{(v)},  \frac{1}{\sigma_m}\,  \nabla \mathcal S_v\bigl(\boldsymbol{\chi}\,\boldsymbol{\psi}_n^{(v)}\bigr)  \right\rangle_{\Omega},  \\
    (\mathsf S_{vw})_{mn}  &=  \left\langle  \boldsymbol{\psi}_m^{(v)},  \frac{1}{\sigma_m}\,  \nabla \mathcal S_v\bigl(\boldsymbol{\chi}\,\boldsymbol{\psi}_n^{(w)}\bigr)  \right\rangle_{\Omega},  \\
    (\mathsf S_{ws})_{mn}  &=  \bigl\langle  \boldsymbol{\psi}_m^{(w)}, \nabla\mathcal S \,\psi_n^{(s)}  \bigr\rangle_{\Omega},  \\
    (\mathsf G_{ww})_{mn}  &=  \bigl\langle  \boldsymbol{\psi}_m^{(w)},  \boldsymbol{\chi}^{-1}\boldsymbol{\psi}_n^{(w)}  \bigr\rangle_{\Omega},  \\
    (\mathsf S_{wv})_{mn}  &=  \left\langle  \boldsymbol{\psi}_m^{(w)},  \frac{1}{\sigma_m}\, \nabla  \mathcal S_v\bigl(\boldsymbol{\chi}\,\boldsymbol{\psi}_n^{(v)}\bigr)  \right\rangle_{\Omega},  \\
    (\mathsf S_{ww})_{mn}  &=  \left\langle  \boldsymbol{\psi}_m^{(w)},  \frac{1}{\sigma_m}\, \nabla\mathcal S_v\bigl(\boldsymbol{\chi}\,\boldsymbol{\psi}_n^{(w)}\bigr)
    \right\rangle_{\Omega},
  \end{align}
\end{subequations}
the elements of the right-hand side are defined as
\begin{subequations}
  \begin{align}
    (\mathsf b_s)_m    &=    -    \frac{1}{\sigma_p}    \bigl\langle    \psi_m^{(s)},    \mathcal D_v^{*}\mathbf J_p    \bigr\rangle_{\Gamma},    \\
    (\mathsf b_v)_m    &=    -    \frac{1}{\sigma_p}    \bigl\langle    \boldsymbol{\psi}_m^{(v)},    \nabla \mathcal S_v\mathbf J_p    \bigr\rangle_{\Omega},    \\
    (\mathsf b_w)_m    &=    -    \frac{1}{\sigma_p}    \bigl\langle    \boldsymbol{\psi}_m^{(w)},    \nabla\mathcal S_v\mathbf J_p    \bigr\rangle_{\Omega},
  \end{align}
\end{subequations}
and where the dependence from $\mathbf r$ is omitted and $\langle a,b\rangle_{\Gamma}  =  \int_{\Gamma} a(\mathbf r)\,b(\mathbf r)\,d\Gamma$ and $\langle \mathbf a,\mathbf b\rangle_{\Omega}  =  \int_{\Omega} \mathbf a(\mathbf r)\cdot \mathbf b(\mathbf r)\,d\Omega$.

The solution $(\mathsf x_s, \mathsf x_v, \mathsf x_w)$ for a given set of primary dipoles forming $J_p$ is then used to evaluate the potential at arbitrary observation points in the whole domain by means of
\begin{equation}
  \phi(\mathbf r)  =  \sum_{n=1}^{N_s} (\mathsf x_s)_n\,(\mathcal S\psi_n^{(s)})(\mathbf r)  -  \sum_{n=1}^{N_v} (\mathsf x_v)_n\,\bigl(\mathcal S_v(\boldsymbol{\chi}\boldsymbol{\psi}_n^{(v)})\bigr)(\mathbf r)  -  \sum_{n=1}^{N_w} (\mathsf x_w)_n\,\bigl(\mathcal S_v(\boldsymbol{\chi}\boldsymbol{\psi}_n^{(w)})\bigr)(\mathbf r)  -  \frac{1}{\sigma_p}\,(\mathcal S_v\mathbf J_p)(\mathbf r).
  \label{eq:curr2phi}
\end{equation}
which can be used to assemble the lead-field matrices.
In this work, a 64-channel montage based on an extended 10-20 system is used.

\subsection*{Anatomical data}

Anatomical information is required for both the data and the visualization module.
In the former, structural features are used to construct the lead-field matrix as presented above; in the visualization module, the same meshes provide the geometrical substrates on which functional data are rendered.
The anatomical surfaces of the scalp, skull, and gray matter are extracted from magnetic resonance imaging (MRI), and more specifically, from the IIT Human Brain Atlas v.5.0~\cite{zhang2018evaluation, qi2021regionconnect}.
A set of meshes is obtained after segmentation and tessellation operations conducted using the simNIBS framework~\cite{puonti_accurate_2020, thielscher_field_2015}.
Then, the surface meshes are subjected to additional manual post-processing in Blender with the goal of cleaning the meshes and of ensuring a closed and nested setting.
In addition, a volumetric mesh suited for volumetric integral equation simulations is obtained using Gmsh~\cite{geuzaine2009gmsh} starting from the inner and outer surface of the skull layer.
The volumetric discretization used for the FEM reference is generated from the same anatomical geometry, although the two formulations rely on a different numerical representation of the electric parameters of the anisotropic media.

\begin{figure}
\centering
\includegraphics[width=0.6\linewidth]{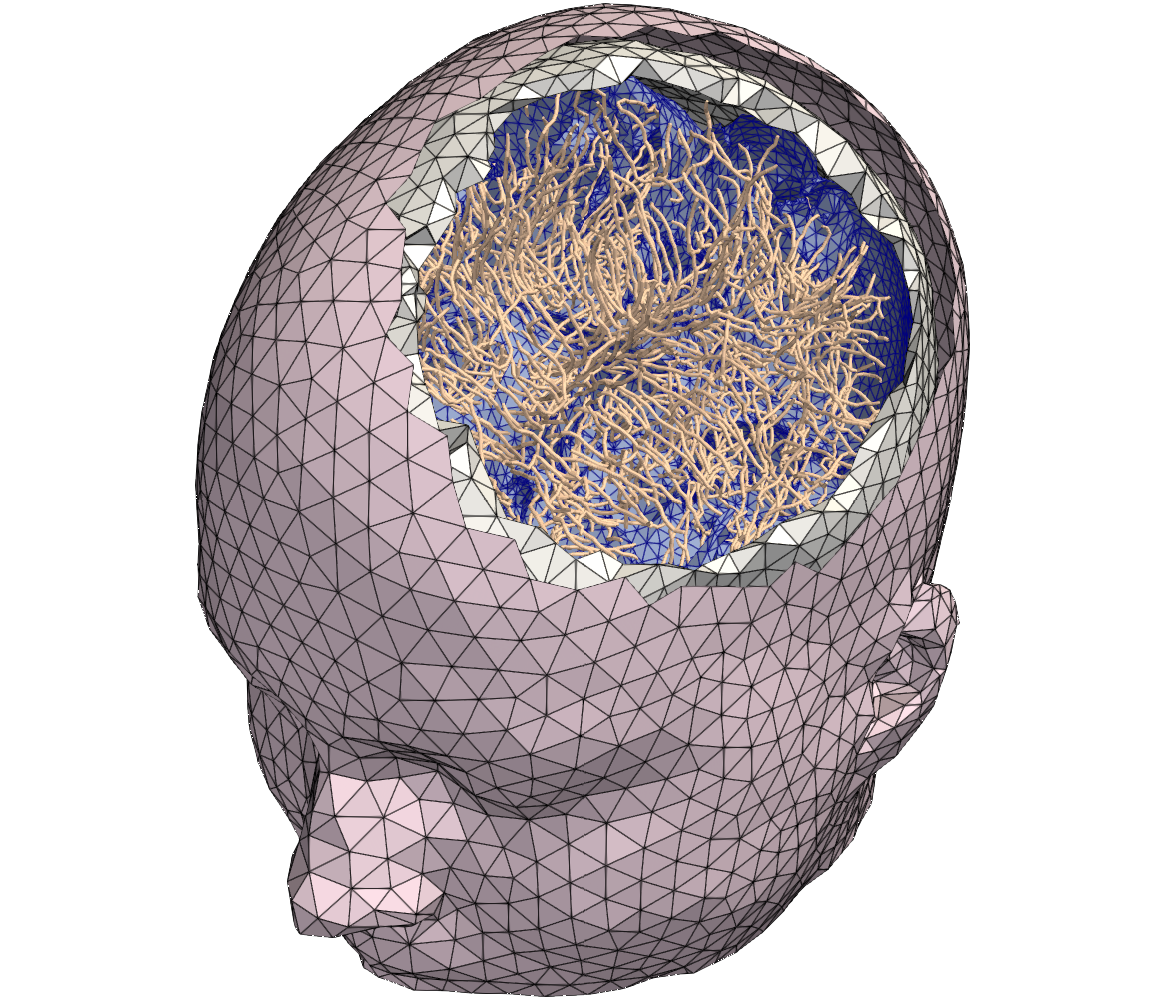}
\caption{Anatomical model used in the proposed framework, including MRI-derived nested surfaces and DTI-derived white matter fiber tracts. Some elements of the outer layers have been hidden to allow the inspection of the inner components.}
\label{fig:model}
\end{figure}

White matter fiber tracts are extracted following a different pipeline.
Starting from DTI data, a set of fibers is derived using a probabilistic tractography method based on second-order integration over fiber orientation distributions~\cite{tournierimproveda, smith2012anatomicallyconstrained}.
Then, to obtain a representation suitable for real-time VR visualization, the streamlines are bundled to reduce their number to approximately 500 fibers and their radius is set to match a realistic volume for white matter~\cite{zhang_universal_2000}.
In this case, a uniform radius for all fibers is considered, but it should be noted that the method allows the definition of a different radius for each fiber, as it is convenient for some methods in the literature~\cite{smith_sift2_2015,daducci_commit_2015}.
The resulting anatomical model, including the nested surface meshes and the bundled white matter fiber representation, is shown in Fig.~\ref{fig:model}.

In addition to the geometrical information, the evaluation of lead-field matrices requires also the characterization of the conductivity of the tissues.
Common conductivity values are used for homogeneous isotropic compartments~\cite{dannhauer2010modeling, vorwerk2014guidelineb}, which are common between the hybrid volume-surface-wire formulation and for the FEM reference.
The skull conductivity, instead, is modeled according to the approach described by Dannhauer et al.~\cite{dannhauer2010modeling}, which mimics the three-layered structure of the skull by imposing radial and tangential conductivity values
\begin{subequations}
  \begin{align}
    \sigma_{\mathrm{rad}}
    &=
    \left(
      \frac{f}{\sigma_{\mathrm{spong}}}
      +
      \frac{1-f}{\sigma_{\mathrm{comp}}}
    \right)^{-1},\\
    \sigma_{\mathrm{tan}}
    &=
    f\,\sigma_{\mathrm{spong}}
    +
    (1-f)\,\sigma_{\mathrm{comp}}
  \end{align}
\end{subequations}
where $\sigma_{\mathrm{spong}}$ and $\sigma_{\mathrm{comp}}$ are the conductivity values of the spongy and compact bones, respectively.
The parameter $f$ models the portion of spongy bone throughout the entire skull and can be extracted from the MRI data.

Finally, the treatment of white matter anisotropy differs between the FEM reference and the hybrid volume–surface–wire formulation.
In the FEM model, anisotropy is represented through a volumetric conductivity tensor obtained by assuming a linear relationship between the water diffusion tensor measured with DTI and the electrical conductivity tensor~\cite{tuch2001conductivityb, tuch1999conductivity, gullmar2010influencea}.
This approach is well-suited for a method leveraging volumetric discretization such as FEM.
In fact, in this case a full tensor can be assigned to each element.
In the hybrid volume-surface-wire formulation, instead, white matter tracts are represented as one-dimensional wire inclusions embedded in the surrounding medium.
Therefore, the DTI-derived volumetric tensor field is replaced by a model that assigns a different conductivity along the local tract direction.
In this model, conductivity is assumed to be equal to the background conductivity in the directions transverse to the fiber and larger along the longitudinal direction, consistently with the common assumption that white matter conductivity is preferentially oriented along fiber pathways~\cite{nicholson_specific_1965}.
The conductivity tensor assigned to a fiber is, then,
\begin{equation}
  \boldsymbol{\sigma}(\mathbf r)  =  \sigma_{iw}\mathbf I  +  \bigl(\sigma_{\ell}-\sigma_{iw}\bigr)  \hat{\boldsymbol{\ell}}(\mathbf r)\hat{\boldsymbol{\ell}}(\mathbf r)^{\mathsf T},  \qquad  \mathbf r \in \Omega_{iw}.
\end{equation}
where $\sigma_{iw}$ is the conductivity of the background medium $\Omega_{iw}$, $\sigma_\ell$ is the longitudinal conductivity, and $\hat{\boldsymbol{\ell}}(\mathbf r)$ is the orientation of the fiber.
Also the choice of $\sigma_{iw}$ and $\sigma_\ell$ is based here on previous studies in the literature~\cite{lozano2014brain}  and kept fixed across simulations.

\subsection*{Data pipeline}

The data module is responsible for EEG acquisition, signal processing, and preparation of the functional data streams sent to the visualization module.
In addition to standard preprocessing operations, such as artifact removal and filtering, the module computes activity at the scalp, cortical, and fiber level from the raw electrode potentials.
Because EEG electrodes provide values only at discrete locations, interpolation is performed to map potentials onto the whole scalp surface.
This operation is performed using spherical splines~\cite{perrin_spherical_1989}.

For cortical sources and fibers, instead, the source imaging problem must be solved first.
Once the lead-field has been defined, a source imaging algorithm (such as sLORETA or wMNE) is used to solve the inverse problem and retrieve the sources.
Afterwards, starting from the estimated dipolar currents, \eqref{eq:curr2phi} can be used to obtain the potentials at the ends of each fiber as well as at arbitrary locations in the head model.
It should be noted that the reconstructed source activity could also be exploited to estimate the potential on the scalp surface, but this operation would require an additional computational overhead compared to interpolation.

Because of the limited coupling between the data and visualization modules, it is possible to adjust few elements of the former before the data are sent to the visualization module to change which data are displayed.
For instance, few changes allow us to visualize the error of the retrieved activity with respect to a reference or the difference between the data retrieved by two different methods in place of visualizing the retrieved activity itself.
This consideration holds for the EEG data used as well.
The data module currently implements a direct connection to an EEG acquisition device during ongoing recordings, the reproduction of previously recorded EEG data, and the usage of simulated EEG signals with known ground-truth source configurations.

\subsection*{Visualization and interaction}

The visualization module receives data divided into three LSL streams produced by the data module: per-vertex values on the scalp mesh, per-vertex values on the cortical mesh, and per-fiber potential differences evaluated at tract endpoints.
At startup, the module binds each stream to its target mesh and initializes the shaders and the user interface controls.
During runtime, each incoming sample is applied to the corresponding vertex attributes or fiber records, and rendered in an immersive scene at the headset refresh rate.
For surfaces, the values are interpreted as scalars per mesh vertex at each time step and passed to custom shaders that apply perceptually uniform sequential colormaps to maximize the interpretability of the data displayed~\cite{borland_rainbow_2007}.
For fibers, the potential difference between endpoints is used to rank tracts at each frame, after which a portion of the stream, which can be controlled by the operator, is marked as active and rendered with an emissive band moving along the fiber.
Alternative activation strategies, such as absolute thresholds, can be enabled with minimal changes but are not explored in this work.

The user interactions follow commonly used gestures.
Users can freely move around and inside the head model and use controllers or hand tracking to reposition, rotate, and scale the model until the desired viewpoint is achieved.
The colormap can be chosen through a menu and layer-specific controls allow the user to toggle visibility and adjust transparency.
In this way, when inspecting the relationship between scalp potentials, cortical sources, and fibers, users can temporarily reduce the opacity of the scalp and skull to reveal the cortex and white matter fiber tracts, then restore them to assess consistency with the observed EEG pattern.
The menus are spatial and follow the same interaction paradigm as the model, reducing cognitive load when switching between tools.
Together, these mechanisms let users fluidly move from a global overview to an analysis in detail and back, keeping attention on the evolving data rather than on the interaction mechanics.

The VR environment was developed in Unity to ensure broad compatibility with head-mounted displays and to simplify future extensions.
This modular design also eases integration of additional data layers or alternative inverse solutions, since only the stream bindings and the shader inputs need to be adapted while the interaction model remains unchanged.

\section*{Conclusion}

In this work, we introduced a hybrid volume-surface-wire formulation in the context of EEG source imaging.
The proposed approach was assessed against an anisotropic FEM reference constructed from the same anatomical geometry, showing overall agreement in the forward solutions and comparable source imaging performance for both wMNE and sLORETA, under both noise-free and noisy conditions.
These results indicate that the hybrid formulation can provide source-imaging results comparable to those obtained with the anisotropic FEM reference without requiring a volumetric discretization of the entire head.
At the same time, the explicit wire representation provides direct access to quantities associated with white matter fiber tracts, which in a FEM framework would instead need to be extracted a posteriori from the volumetric solution.
Building on this property, we developed a real-time visualization tool that integrates scalp potentials, reconstructed cortical activity, and tract-related quantities, allowing the computed data to be visualized and explored interactively

\section*{Data availability}
The datasets generated and analyzed in this study are publicly available in the Zenodo repository (DOI: \href{https://doi.org/10.5281/zenodo.22042624}{10.5281/zenodo.22042624}. The data include the lead-field matrices computed using the hybrid volume–surface–wire formulation, the anisotropic FEM reference model, and the source-jittered anisotropic FEM model used to generate the synthetic EEG data. The anatomical data used to construct the head models were obtained from the publicly available IIT Human Brain Atlas v.5.0.

\bibliography{2025_Neurosurf.bib}


\section*{Acknowledgment}
This work has received funding from the European Research Council (ERC) through the HORIZON ERC Proof of Concept Grants under Grant 101189419 (Project TurboEEG).

\section*{Author contributions statement}

P.R. contributed to the conception and design of the work, developed the methodology and software, performed the analysis, interpreted the results, and drafted the manuscript; M.M. and A. Ma. contributed to the creation and development of software used in the work; A. Me. contributed to the analysis and interpretation of the results; F.P.A. contributed to the conceptualization of the work and supervised the study. All authors reviewed and approved the final manuscript.

\section*{Competing interests}
The authors declare no competing interests.



\end{document}